\documentclass[a4paper]{article}

\usepackage{icrc2013}

\title{Sagittarius dwarf spheroidal galaxy observed by H.E.S.S. }

\shorttitle{Sagittarius dwarf spheroidal galaxy observed by H.E.S.S. }

\authors{
G. Lamanna$^{1}$,
C. Farnier$^{2}$,
A. Jacholkowska$^{3}$,
M. Kieffer$^{3}$,
C. Trichard$^{1}$
for the H.E.S.S. Collaboration.
}

\afiliations{
$^1$ LAPP, Universit\'{e} de Savoie, CNRS/IN2P3, F-74941 Annecy-le-Vieux, France. \\
$^2$ OKC, Physics Department, Stockholm University, AlbaNova SE-10691 Stockholm, Sweden. \\
$^3$ LPNHE, Universit\'{e} Paris VI et Paris VII, CNRS/IN2P3, 4 Place Jussieu, F-75252, Paris Cedex 5, France. \\

}

\email{Giovanni.Lamanna@lapp.in2p3.fr}

\abstract{Dwarf spheroidal galaxies are characterized by a large measured mass-to-light ratio and are not expected to be the site of high-luminosity non-thermal high-energy gamma-ray emissions. Therefore they are among the most promising candidates for indirect searches of dark matter particle annihilation signals in gamma rays. The Sagittarius dwarf spheroidal galaxy has been regularly observed by the High Energy Stereoscopic System (H.E.S.S.) of Cherenkov telescopes for more than 90 hours, searching for TeV gamma-ray emission from annihilation of dark matter particles. In absence of a significant signal, new constraints on the annihilation cross-section of the dark matter particles applicable for Majorana Weakly Interacting Massive Particles (WIMPs) are derived.}

\keywords{dark matter, gamma rays, H.E.S.S., dwarf galaxy}

\begin{document}
\maketitle


\section{Searching for dark matter}

A large number of observations from Galactic to cosmological scales support the explanation that Dark Matter $\mbox{(DM)}$ would be composed by a new type of particle although its nature remains unknown. The proposal of WIMP predicted by theories beyond the Standard Model of particle $\mbox{physics}$ provides a relic abundance accounting for the inferred amount of DM \cite{bib:[2]}.
The WIMP search is conducted in three ways: by particle production at the LHC, probing the theories of Standard Model extension; by searching for nuclear recoil signals experiments, probing the WIMP scattering cross section; by indirect searches of a signal in the secondary products of WIMP annihilation, probing the corresponding cross section.

Potential spectral signatures in gamma rays can be classified in mainly strong spectral features and ambiguous signals. The first class is for instance due to annihilation into $\gamma$$\gamma$ or $Z$$\gamma$ producing a sharp line spectrum with a photon energy depending on the WIMP mass, e.g. the Supersymmetric (SUSY) neutralino hypothesis. Unfortunately, these processes are loop-suppressed and therefore very rare. To some extent more ambiguous are signals due to continuum emission from pion decay resulting from the WIMPs annihilation in pairs of leptons or quarks. The number of gamma rays finally originated by WIMP annihilation depends quadratically on the DM density along the line of sight of the observer. This motivates a number of promising targets for indirect DM searches, namely those with known density enhancements, in particular the Galactic Centre and close-by dwarf galaxies and galaxy clusters. 
More specifically, assuming the $\Lambda$CDM cosmological model, the hierarchical collapse of small over-densities are formed by DM structures which may also host smaller satellite structures and it has been proposed that dwarf spheroidal galaxies may have formed within some of these sub-halos hosted in the larger Milky Way DM halo \cite{bib:[5]}.

\section{Sagittarius Dwarf Spheroidal Galaxy}

The Sagittarius Dwarf Spheroidal Galaxy (SgrD) was discovered in 1994 and it is located at RA = 18$^{\mathrm{h}}$ 55$^{\mathrm{m}}$ 59.9$^{\mathrm{s}}$, Dec = -30$^{\circ}$ 28$^{\mathrm{\prime}}$ 59.9$^{\mathrm{\prime}\mathrm{\prime}}$ in equatorial coordinates (J2000.0), at a distance of about 24 kpc from the Sun \cite{bib:[4]}. SgrD is one of the nearby dwarf spheroidal companion galaxies of the Milky Way. Characterised by an extremely low surface brightness it has been estimated that SgrD orbits the Milky Way within less than a billion years and therefore it must have already passed through the galactic plane at least about ten times. Although its high level of tidal disruption, the hypothesis of a large DM content would be responsible to hold onto so many of its stars and for so long.

 \begin{figure*}[t]
  \centering
  \includegraphics[width=0.67\textwidth]{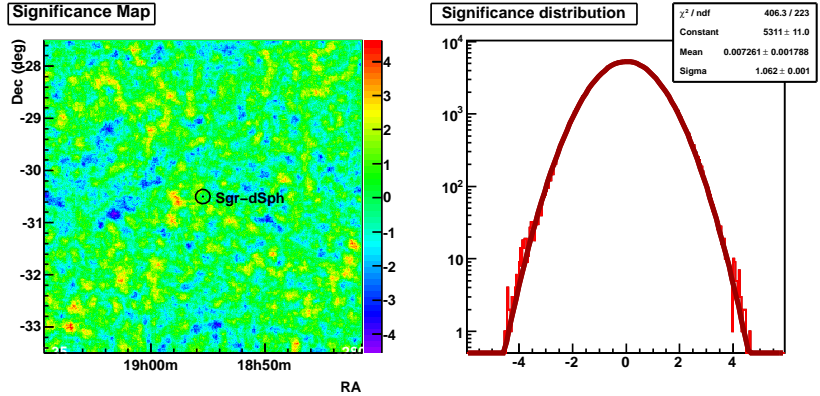}
  \caption{Significance sky map: no excess observed at the target position (left panel). Significance distributions related to the on-source and off-source events which are well overlapped and compatible each other (right panel).}
  \label{fig:sigmas}
 \end{figure*}

\section{H.E.S.S.}

The High Energy Stereoscopic System (H.E.S.S.) is an array of Imaging Atmospheric Cherenkov Telescopes (IACTs) the working principle of which is based on the detection of faint Cherenkov light from the gamma-ray-induced air showers in the atmosphere. Located in the Khomas Highland in Namibia, 1800 m above sea level, H.E.S.S. began operation in 2003 with four 13 m dish-diameter telescopes, equipped with cameras containing 960 photo-multiplier tubes (PMTs) and operated in a coincidence mode, in which at least two of them must have been triggered for each event within a coincidence window of 60 ns. Since September 2012 one additional larger 28 m dish-diameter telescope (and a camera with 2048 PMTs) has started operation enabling to achieve a larger sensitivity of the full array in the range of about 100 GeV to 10 TeV while increasing the lower energy domain to $\sim$ 30 GeV by exploiting the larger telescope in standalone mode.  

The H.E.S.S. collaboration observed SgrD searching for a potential emission of very-high-energy (VHE) gamma rays during 11 h in 2006. In absence of any signal an upper limit on the gamma-ray flux and a constraint on the velocity-weighted annihilation cross-section in the frame of SUSY models as well as extra-dimensions models were published \cite{bib:para}. A deeper observation of SgrD was conducted accumulating further 90 hours of observation in 2007-2012 using the four 12 m dish telescopes. In the following the procedure and the results of the analysis of these new observations are described and discussed.

\section{Observations and data analysis}

H.E.S.S. has observed SgrD from 2006 to 2012. The observations were performed in wobble mode, i.e. with the target offset by 0.7$^{\circ}$ to 1.1$^{\circ}$ from the pointing direction, enabling simultaneous background estimation and in the same field of view. The data used for the analysis were taken at different zenith angles spanning from few degrees up to 45$^{\circ}$ and an average value of $\sim$ 14$^{\circ}$ over the full data sample. Some standard quality selection cuts were applied resulting in a reduction of the data of about 13\% corresponding to a total 91.5 live-hours data analysed. 

The $X_{eff}$ analysis was employed for the selection
of gamma-ray events and for the suppression of cosmic-ray
background events. $X_{eff}$ denotes a multivariate analysis
method developed to improve signal-to-background discrimination, which
is important in searches for weak signals \cite{bib:xeff}. The $X_{eff}$
method improves the separation of gamma and cosmic-ray events
compared to the standard H.E.S.S. analysis \cite{bib:aha06b}, by exploiting the complementary discriminating
variables of three reconstruction methods\footnote{The three reconstruction methods are referred to as Hillas \cite{bib:hillas}, Model \cite{bib:model} and 3D-model \cite{bib:model3D, bib:godo}.} in use in the H.E.S.S. analysis. The resulting unique discriminating variable $X_{eff}$ acts as an event-by-event gamma-misidentification probability estimator. The definition of the $X_{eff}$ probability function follows the relation:
\begin{equation}
X_{eff}(d_i) = \frac{{\eta} {\displaystyle\prod_{j}H_j(d_i)}}{(1-{\eta}){\displaystyle\prod_{j}G_j(d_i)+{\eta}\prod_{j}H_j(d_i)}},
\label{global}
\end{equation}
\noindent
where $d_i$ are the discriminating variables (indexed by $i$) of three reconstruction methods (indexed by $j$); $G_j(d_i)$ and $H_j(d_i)$ are the one-dimensional {\it probability density functions} ({\it p.d.f.}s) for events identified as gamma-ray-like ($G$) and hadron-like ($H$) (the product of individual {\it p.d.f.}s replaces the global multi-dimensional ones since the variables $d_i$ are highly uncorrelated, for gamma-ray events in particular); $\eta$ is the misidentified fraction of the $gamma$ class of events (i.e. the relative background fraction). The final gamma-ray event selection was achieved with the set of cuts adapted to the detection of faint sources ($\eta=0.7$, $X_{eff,cut}=0.3$) and a cut in the reconstructed image charge requiring more than 60 photo-electrons. The gamma-ray signal was searched at the target position within an angular size of $\theta^2$ $\leq$ 0.02$^{{\circ}2}$.

\section{Results}

\begin{figure}[!h]
  \centering
  \includegraphics[width=0.43\textwidth]{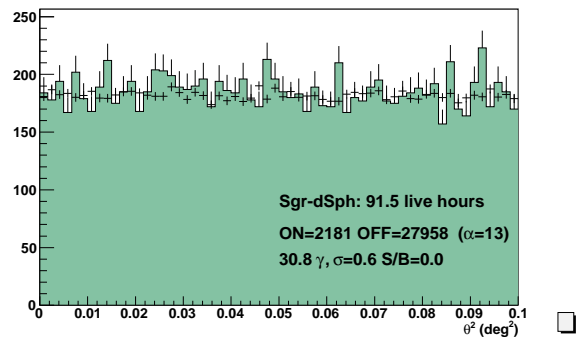}
  \caption{$\theta^2$ radial distribution of the ON events for gamma-ray-like events from the SgrD target position. Estimated background is also shown by black markers overlapped to the histogram. No excess is seen at small $\theta$ value.}
  \label{fig:theta}
 \end{figure}

No significant gamma-ray excess was found above the estimated background at the nominal position of SgrD nor in the camera field of view as shown by the significance sky-map and the Gaussian distribution compatible with the background in figure \ref{fig:sigmas} as well as presented by the flat $\theta^2$ distribution in figure \ref{fig:theta}. The target position is chosen according to the photometric measurements of the SgrD luminous cusp showing that the position of the centre corresponds to the centre of the globular cluster M54 \cite{bib:[17]}: RA = 18$^{\mathrm{h}}$ 55$^{\mathrm{m}}$ 59.9$^{\mathrm{s}}$, Dec = -30$^{\circ}$ 28$^{\mathrm{\prime}}$ 59.9$^{\mathrm{\prime}\mathrm{\prime}}$ in equatorial coordinates (J2000.0).

A 95\% confidence level (C.L.) upper limit (N$_{\gamma}^{95\% C.L.}$) on the total numbers of observed gamma-ray events
 can be thus deduced using the Feldman \& Cousins \cite{bib:[18]} method. This limit is computed using the number of events registered on target region ($ON$=2181), those corresponding to the background ($OFF$=27958) normalized by the ratio ($\alpha$=13) between the on-source and the off-source livetimes. The result is N$_{\gamma}^{95\% C.L.}$ = 123. Effects of uncertainties on background and $\alpha$ can be safely neglected. 

Such a limit can be used to constrain indirectly the WIMP annihilation cross-section in a specific modeling context. In order to do so it is worth to remind that the differential gamma-ray flux ($d\Phi_\gamma/dE_\gamma$) due to DM particle annihilation depends on the following items:
\begin{itemize}
\item the particle physics coupling of the DM particle;
\item the intrinsic DM density distribution in the source;
\item the field of view $\Delta\Omega$ within which the signal is integrated along the line of sight by the observer.
\end{itemize}
\noindent
It is usually factorised into two terms:

\begin{equation}
\frac{d\Phi_\gamma}{dE_\gamma}(E_\gamma, \Delta\Omega)  =  \Phi^{pp}(E_\gamma) \times J(\Delta\Omega)\Delta\Omega,
\end{equation}

where the first factor ($\Phi^{pp}$) takes into account the particle physics model for the WIMP annihilation, while the second factor (called $J$-factor in the following) accounts for the astrophysics model and denotes the DM density distribution in the source. 

The particle physics factor is given by:

\begin{equation}
\Phi^{pp} = \frac{d\Phi_\gamma}{dE_\gamma}  =  \frac{1}{8\pi} \frac{\langle\sigma_{ann}v\rangle}{m^2_\chi}  \times  \frac{dN_\gamma}{dE_\gamma},
\end{equation}
\noindent
where $\langle\sigma_{ann}v\rangle$ is the velocity-averaged annihilation cross section, $m_\chi$ is the WIMP particle mass, and $dN_\gamma/dE_\gamma$ is the differential gamma-ray spectrum summed over the whole final states weighted by their corresponding branching ratios. In this work, for simplicity and homogeneity with previous similar estimates, all channels are taken into account by using a  parametrised average spectrum (from Bergstr$\mathrm{\ddot{o}}$m et al. \cite{bib:berg}):

\begin{equation}
\frac{dN_\gamma}{dE_\gamma}  =  \frac{1}{ m_\chi } \frac{ dN_\gamma }{dx} = \frac{1}{ m_\chi } \frac{0.73 e^{-7.8x}}{x^{1.5}},
\end{equation}
\noindent
where $x$ = $E_\gamma/m_\chi$.

In the astrophysical factor, the integral along the line of sight of the squared density of the DM
distribution in the object is averaged over the instrument solid angle of the integration region. For this analysis with
H.E.S.S., $\Delta\Omega$ = 2 $\times$ 10$^{-5}$ sr since we are looking for an almost point-like signal considering the point-spread function of the instrument:
\begin{equation}
J = \frac{1}{\Delta\Omega}\int\limits_{\Delta\Omega}^{ }\int \rho^2_{DM}(l,\Omega)dld\Omega.
\label{j}
\end{equation}

Two different models of the DM halo of SgrD were considered in the analysis published by the H.E.S.S. collaboration in 2008 \cite{bib:para}. They turned out to be too optimistic since the dwarf disruption by tidal winds was ignored. Recent measurements (by Niederste-Ostholt et al. \cite{bib:nied}) have made possible to scale down the previous density estimate since only 30-to-50\% of the luminosity of SgrD is assumed currently still bound to the remnant core. Recently new modeling of the SgrD density which takes into account tides have been published \cite{bib:[16]}. Such new densities are considered in this work. For completeness the approach followed and the results obtained in \cite{bib:[16]} are here summarised.
\begin{table}[!b]
\begin{center}
\begin{tabular}{|c|c|c|}
\hline Dark Matter density profile & $J$-factor (GeV$^2$ cm$^{-5}$) \\ \hline
 Isothermal   & 0.88$\times$10$^{23}$ \\ \hline
 NFW   & 1.00$\times$10$^{23}$ \\ \hline
\end{tabular}
\caption{Astrophysics J-factor computed for two different dark matter halo profi
les~\cite{bib:[16]}.}
\label{table:j}
\end{center}
\end{table}
A DM halo described by NFW profile \cite{bib:[15]} is first considered:
\begin{equation}
\rho_{NFW}(r) = \frac{\rho_s}{(r/r_s)(1+r/r_s)^2},
\end{equation}
\noindent
where parameter values: $r_s$ = 1.3 kpc is a scale radius and $\rho_s$ = 1.1 $\times$ 10$^{−2}$M$_{\odot}$ pc$^{−3}$ is a characteristic density.

An isothermal profile for DM halo is also considered. It is proposed by Pe$\mathrm{\tilde{n}}$arrubia et al. \cite{bib:[13]} assuming that SgrD is a late-type, rotating disc galaxy. In this model, the galaxy is composed of a stellar disk and a DM component with the following density distribution:
\begin{equation}
\rho_{ISO}(r) = \frac{m_h\alpha}{2\pi^{3/2}r_{cut}} \frac{exp[-(r/r_{cut})^2]}{(r^2_c + r^2)},
\end{equation}
\noindent
where $m_h$ is the halo mass, $r_c$ is the core radius and $\alpha$ $\simeq$ 1.156. It is found that the properties of the
stream are not particularly sensitive to the value of the choice of the core radius at $r_c$ = 0.45 kpc. Assuming an initial luminosity of $\sim$ 10$^8$ L$_{\odot}$ \cite{bib:nied}  and a mass-to-light ratio of 24 \cite{bib:[14]}, the total mass of the halo is found to be $m_h$ = 2.4 $\times$ 10$^9$M$_{\odot}$. To take into account the lost of the outer halo envelope due to tidal disruption by the Milky Way, a truncation in the DM density profile is imposed at $r_{cut}$ = 12 $r_c$ = 5.4 kpc. It roughly corresponds to the tidal radius of a satellite galaxy at pericenter of 15 kpc with a mass $\sim$ 3 $\times$ 10$^9$M$_{\odot}$.

The $J$-factors can be computed using the equation \ref{j}. They are summarised for both approaches in table \ref{table:j} as extracted from \cite{bib:[16]}. It is important to remind that according to the authors of \cite{bib:[16]} the uncertainties on the halo profile parameters are still large and can affect the value of the astrophysical factor by a factor of 2.
 \begin{figure*}[t]
  \centering
  \includegraphics[width=0.56\textwidth]{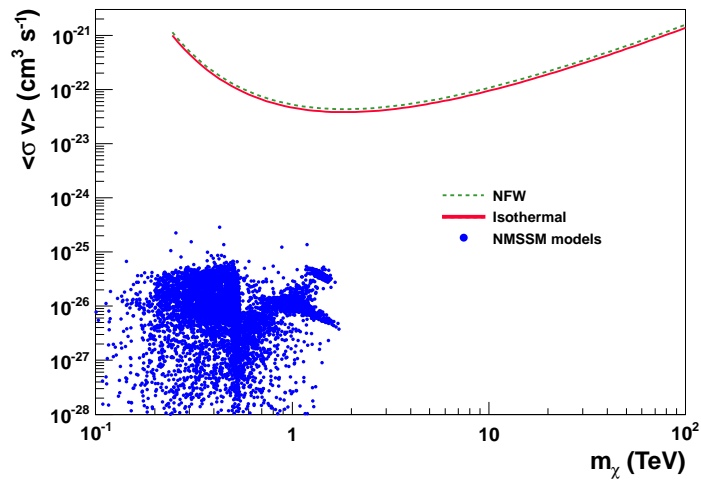}
  \caption{Exclusion limit at 95\% C.L. on the velocity-weighted annihilation cross-section versus the DM particle mass for a NFW (green dotted line) and an Isothermal (red line) DM density profiles. NMSSM models scan is shown (blue markers).}
  \label{fig:limit}
 \end{figure*}
Finally the 95\% C.L. upper limit on the velocity-weighted annihilation cross-section as a function of the WIMP particle mass and for a given halo profile is computed as:

\begin{equation}
\langle\sigma v\rangle^{95\% C.L.}_{min} = \frac{8\pi}{J(\Delta\Omega) \Delta\Omega } \times \frac{m^2_{DM} N_{\gamma, tot}^{95\% C. L.}}{T_{obs} \int\limits_0^{m_{DM}} A_{eff} (E_\gamma) \frac{dN_\gamma}{dE_\gamma} (E_\gamma)dE_\gamma},
\end{equation}
\noindent
where $T_{obs}$ is the observation time dedicated to SgrD (91.5 hours) and $A_{eff}(E)$ is the H.E.S.S. effective area. The exclusion curves for the SUSY neutralino are shown in figure \ref{fig:limit} referring to the two halo profiles as in table \ref{table:j}. The exclusion limits depend on the particle  mass and the best sensitivity is reached at 1-2 TeV with the value of $\sim$ 4$\times$10$^{-23}$ cm$^3$ s$^{-1}$ still well above the thermal value of $\langle\sigma v\rangle$ $\sim$ 3$\times$10$^{-26}$ cm$^3$ s$^{-1}$ \cite{bib:[3]}. As the two astrophysical $J$-factors are compatible one each other it results that the two limits are of the same order of magnitude.

In order to compare the H.E.S.S. exclusion limit to the predictions of a set of realistic particle physics models, we performed a scan of the Next-to-Minimal Supersymmetric Standard Model (NMSSM)\cite{bib:nmssm} that constitutes one of the most well-motivated extensions of the Standard Model, both theoretically and phenomenologically, especially in light of the recent discovery of a Higgs-like particle at the LHC \cite{Arbey:2011ab,Arbey:2011ab2}. The scan was performed\footnote{Study conducted within the French ANR-$DMAstroLHC$ project by A. Goudelis, P.D. Serpico and G. Lamanna} using the publicly available code NMSSMTools 3.2.4 \cite{Ellwanger:2005dv,Das:2011dg,Muhlleitner:2003vg} with parameter values and ranges as: $\tan$$\beta$ $\in$ [1.5, 60]; $M_1$, $M_2$ $\in$ [100, 2000]$\rm{GeV}$; $M_3$ $\in$ [500, 3000]$\rm{GeV}$; $\lambda$, $\kappa$ $\in$ [0.1, 0.8]; $A_\kappa$ $\in$ [-2000, 0]; $\mu_{\rm{eff}}$ $\in$ [100, 2000]; $M_A$ $\in$ [10, 2000] and making several discrete choices for $A_{U_3}$, $A_{D_3}$ = $A_{E_3}$, $m_{L_3}$ = $m_{E_3}$ = $m_{Q_3}$, $m_{U_3}$ and $m_{D_3}$ depending on the scanned subregion. Models with a Higgs particle in the mass range [119; 130]$\rm{GeV}$ were retained. For computing of the predicted relic density according to WMAP-7 and thermally averaged self-annihilation cross-sections at zero velocity for each considered model, the micrOMEGAs-3.1~\cite{Belanger:2013gh}  was used imposing a loose relic density constraint of $\Omega_{\rm{DM}}$$h^2$ $\in$ [0.087, 0.14], thus resulting in more than 9000 viable points. As a conclusion from this scan, the capability to provide a complementary constraint to LHC seems to be limited for the somewhat less conventional models as inferred from figure 3, however in agreement with mostly studied constraint MSSM models. It would be not necessarily the case for alternative models, e.g. leptophilic models with light mediators~\cite{bib:sommer}, to be studied further.

\section{Conclusions} 
The Sagittarius dwarf spheroidal galaxy has been observed with H.E.S.S. for more than 90 hours, thus enabling efficient search for a dark matter signal coming from this Milky Way satellite. The absence of a signal leads to constraints on the velocity weighted mean annihilation cross-section as a function of WIMP mass which were compared to NMSSM models. The dark matter halo modeling has crucial impact on the derived limits; an increase in detector sensitivity is mandatory to improve the search potential (as expected with the Cherenkov Telescope Array).

\end{document}